\documentclass[a4paper,11pt]{article}
\usepackage[]{latexsym}

\newcommand{\be}{\begin{equation}}
\newcommand{\ee}{\end{equation}}
\newcommand{\bea}{\begin{eqnarray}}
\newcommand{\eea}{\end{eqnarray}}
\renewcommand{\Im}{{\rm{Im}}}

\title{The universal $\ln^{2}s$ increase in total cross sections}
\author{Kyungsik Kang$^{a,b}$ and Bruce H. J. McKellar$^{a,c}$\\
$^{a}$Korean Institute for Advanced Study, \\207-43 Cheonryangni-dong, 
Dongaemun-du, Seoul 130-012 Korea\\
$^{b}$Physics Department, Brown University, Providence, RI 02912, 
USA.\thanks{permanent address.  e-mail: kang@het.brown.edu}\\
$^{c}$School of Physics, University of Melbourne, Victoria Australia 
3010.\thanks{permanent address.  e-mail: 
b.mckellar@physics.unimelb.edu.au}}
\date{February 12, 2003}

\begin{document}

\maketitle

\begin{flushright}
    \vspace{-8.5cm}
    KIAS-P02043
    
    UM-P-017-2002
    \vspace{9cm}
\end{flushright}

\begin{abstract}
    
    While it has long been known that many models of high energy 
    scattering give cross sections which rise as $\ln^{2}s$, the 
    determination of the coefficient of this term is rarely given.  We 
   show that in gaussian and exponential eikonal   models an exact 
   expression for the cross section can be obtained, which 
   demonstrates the $\ln^{2}s$ asyptoptic behaviour and determines its 
   coefficient.  The coefficient is universal, as found empirically, 
   and the value of the constant obtained from the gaussian model is 
   in good agreement with the empirical value.
   
\end{abstract}

\section{Introduction}
Recent analyses of high energy  forward scattering 
data  \cite{HEA} have confirmed that the 
cross sections rise as $\ln^{2} s$, a behaviour first suggested by 
Heisenberg \cite{HEI}, and proportional to  \emph{Froissart-Martin 
Bound} \cite{FMB}.  Similar behaviour has been suggested by Giddings 
\cite{GID} 
in non conformal gauge/gravity duality.  He finds the high energy 
scattering cross section is $\sigma = M^{-2}\ln^{2}s$, with $M$ the 
mass of the lightest Kaluza-Klein excitation.
The coefficient $B$ of the $\ln^{2}s$ term in the fit to the 
cross section is found to be independent of the hadronic reaction 
studied ($\bar{p}p, pp, \Sigma^{-}p, \pi^{\pm}p$, 
and $K^{\pm}p$),
while for $\gamma p$ and $\gamma \gamma$, the the equivalent
coefficents are written as $B_{\gamma} = \delta B$ and $B_{\gamma
\gamma} = \delta ^{2} B$ respectively with the same constant $\delta$
as expected from factorization of the $\gamma$ cross sections.
From all of these cross-sections the  value $B = 0.313(9)\rm{mb}$ is 
determined.  Note that this constant is 
two orders of magnitude smaller than the Lukaszuk-Martin limiting 
value, 
\cite{LMC}
\be
B \le \frac{\pi}{m_{\pi}^{2}} = 62 \rm{mb}.
\ee

This observation raises the question of identifying the physics 
responsible for the value of $B$.

Dosch, Gauron and Nicolescu \cite{DGN} have recently proposed that 
Heisenberg's original model can be extended to obtain the value of 
$B \approx \pi/(4M^{2})$, where $M$ is a glueball mass.

It has long been recognised \cite{FFKT, JEN} that the eikonal
representation of the total cross section can lead to cross sections
rising as $\ln^{2}s$.  In this paper we study three models for the
eikonal and determine the coefficient $B$ for these models.

The three models for the eikonal studied are
\begin{enumerate}
    \item \label{EXP} a factorisable exponential model, motivated by
    meson exchange \cite{FFKT}, 
    \item \label{Gauss} a non-factorisable
    gaussian model, motivated by multiperiphal models \cite{FFKT},
    \item \label{ggq} the QCD motivated model of Block and co-workers
    \cite{MARG,Block}.
\end{enumerate}    

We are able to obtain  exact results for the cross section 
for cases \ref{EXP} and 
\ref{Gauss}, in terms of  incomplete gamma functions and their 
derivatives.  Examination of the asymptotic behaviour gives the 
expected $\ln^{2}s $ rise at large $s$, and determines the 
coefficient $B$ in each case.

For case \ref{ggq}, we have been unable to find an exact solution in 
terms of known functions, but an approximate evaluation of the cross 
section gives $\ln^{2}s$ behaviour, and determines its 
coefficient.

\section{The eikonal representation of the total cross-section}

In terms of the eikonal function $\chi(b,s)$ the scattering amplidude
is represented as 
\be f(s,t) = \imath \int_{0}^{\infty} b db
J_{0}(b\sqrt{-t})\left(1 - e^{-\chi(b,s)}\right) 
\ee 
and the total
cross-section is thus 
\be \sigma_{t} = 4\pi \Im \,f(s,0) = 4\pi
 \int_{0}^{\infty} b db \left(1 -
e^{-\chi(b,s)}\right), 
\ee
where the eikonal is approximated as 
purely real 
corresponding to the black-disc collision at high energies.

Strictly, for $pp$ and $\bar{p}p$ scattering we should introduce
eikonals $\chi_{\pm}(b,s) = \frac{1}{2}\left(\chi_{pp}(b,s) \pm
\chi_{\bar{p}p}(b,s)\right)$ to write crossing-even and crossing-odd
forward amplitudes.  For other processes similar $\chi_{\pm}(b,s)$
eikonals are constructed.  At high energies, we may neglect the
$\chi_{-}$ contribution, and so we use $\chi$ to simply mean
$\chi_{+}$, the crossing-even amplitude.  Physically $\chi (b,s)$ is
made of all elementary inelastic collision cross section of
parton-pairs $\sigma_{0}$ and overlaps of convoluted structure
functions of parton pairs at impact parameter $b$.  Following the
normalization conventions used in mini-jet models\cite{MARG,Block}, we
parametrize the eikonal such that
\be
\int \chi (b, s) d^{2} {\bf b} = \sigma _{0} (s)
\ee
In cases where $\chi(b,s)$ is 
written as the product of a function of $b$ and a function of $s$, the 
eikonal is said to be factorisable, as it is in the Chou-Yang model 
\cite{CY}.

Our technique for evaluating the integral is most straightforward for 
the gaussian case 
which we consider first in the following.

\subsection{The Non-factorisable Gaussian Eikonal}
The gaussian representation of the eikonal is motivated by the 
multi-Pomeron representation of multiperipheral ladder exchanges 
\cite{FFKT}.  One writes
\be
\chi(b,s) = \frac{\pi \lambda} {B(s)} s^{\Delta}
\exp\left\{-\frac{b^{2}}{B(s)}\right\}.
\ee
The parameters $\Delta$ and $B(s)$ are determined by the Pomeron 
trajectory
$\alpha_{P}(t) = \alpha_{P}(0) + \alpha^{\prime}_{P}(0) t,
$
consistent with shrinkage of the forward peak in differential cross sections,
\be
\Delta = \alpha_{P}(0) - 1 \quad\quad \rm{and} \quad \quad B(s) = 2 
\alpha^{\prime}_{P}(0) \ln s + k
\ee
where $k$ is a constant\footnote{There is an unfortunate conflict in 
notation between reference \cite{HEA}, which uses $B$ for the 
coefficient of $\ln^{2}s$, and reference \cite{FFKT}, which uses 
$B(s)$ as the gaussian scale parameter.  We follow this usage, 
warning the reader to note carefully whether $B$ is a constant or a 
function of $s$}.  The logarithmic dependence of $B(s)$ leads to a 
typical shrinkage of the forward peak in the differential 
cross-section.

 The use of $\chi$ as the variable of integration \cite{LL} 
enables us to write the exact result for the total cross-section:
\be
\sigma_{t}(s) = 2\pi B(s) \left(\ln C(s) + \gamma + E_{1}(C(s)) \right)
\ee
where $C(s) = \left(\pi \lambda/B(s)\right) s^{\Delta} = \chi(b=0, s)$ is the 
opacity of the eikonal \cite{Block-2},  
$\gamma$ is Euler's 
constant, and $E_{1}(x) = \int_{x}^{\infty} e^{-t} dt/t$ is the 
exponential integral \cite{EXPINT}.

 For large $x$, $E_{1}(x) \sim \left(e^{-x}/x\right)(1+O(x^{-1})$, and 
 so the large energy behaviour of the cross-section is
 \be
 \sigma_{t}(s) \sim 2\pi B(s)\left(\ln C(s) + \gamma\right),
 \ee
and the leading term in this gives
\be
\sigma_{t}(s) = 4 \pi \alpha^{\prime}_{P}(0) \Delta \ln^{2}s + O(\ln 
s), 
\ee
defining the constant $B$ in terms of the properties of the Pomeron.  
Clearly $B$ is a universal constant.  With $\alpha^{\prime}_{P} = 
0.25 \rm{GeV}^{-2}$ \cite{DONLAN} and with the value
$\Delta$ in the range$(0.1,  0.4)$, determined from deep inelastic scattering 
data \cite{DELDIS}, we obtain
\be
B \in (0.12, 0.48) \rm{mb},
\ee
values which compare favourably with the observed value of $0.31 \rm{mb}$.

This method also gives the $\ln s$ and the constant terms in the total 
cross section, but the coefficients of those terms are not  
process independent, so we do not quote them here.

\subsection{The factorizable exponential eikonal}
The eikonal in this case takes the form 
\be \chi(b,s) =
\frac{\lambda}{2 \pi b_{0}^{2} } s^{\Delta} \exp\left\{ -
\frac{b}{b_{0}}\right\}, 
\ee
and corresponds to massive particle exchange, with $b_{0}$ being 
determined by the mass of the lightest exchanged meson.
One can again use the eikonal as the variable of integration in the 
eikonal expression for the total cross section, which then becomes 
exactly
\bea
\sigma_{t}(s) & = &  2 \pi b_{0}^{2} \Bigg(\ln^{2} C(s) + 
\Bigg[\frac{\pi^{2}}{6} + \gamma^{2} + 2\ln C(s) \gamma\Bigg] 
\nonumber \\
&& -   \left[\Gamma_{1}(0,C(s)) - 2 \ln
C(s) \Gamma(0,C(s))\right]\Bigg), 
\eea
where 
$C(s) = \frac{\lambda}{2\pi b_{0}^{2}} s^{\Delta}$, 
$\Gamma(\alpha, x) =
\int_{x}^{\infty} t^{\alpha - 1} e^{-t} dt$ is the upper incomplete
gamma function, and $\Gamma_{1}(\alpha, s)$ is its partial derivative with 
respect to $\alpha$.

The terms involving $\Gamma_{1}(0,C(s))$ and $\Gamma(0, C(s))$ 
are exponentially small for large $s$, and the high energy behaviour 
is again dominated by the $\ln^{2}s$ term
\be
\sigma_{t}(s) = 2 \pi b_{0}^{2} \Delta^{2} \ln^{2}s + O(\ln s)
\ee
with the coefficient $B = 2 \pi b_{0}^{2}$.  In this case 
$b_{0}^{-1}$ is the mass of the exchanged meson, which can be 
universally taken to be the lightest glueball.  Then with 
$b_{0}^{-1} = 1.4GeV$, and the previous range of values for 
$\Delta$, $B \in (0.006, 0.1) \rm{mb}$, rather smaller than the 
observed value of the coefficient, but, for the larger values of 
$\Delta$, or $1/b_{0}$, or both,  the same order of magnitude for the 
coefficient can readliy be obtained.

\subsection{The QCD inspired eikonal}

Block \emph{et al} \cite{Block} have introduced the eikonal
\be
\chi = \sigma_{gg}(s) W(b;\mu_{gg}) + \sigma_{qg}W(b;\mu_{qg}) + 
\sigma_{qq}W(b;\mu_{qq})
\ee
where $\sigma_{gg}, \sigma_{qg}$ and $\sigma_{qq}$ represent
gluon-gluon, quark-gluon and quark-quark interaction cross sections,
and $W(b;\mu)$ are overlap functions normalised so that $2\pi
\int_{0}^{\infty} b db W(b;\mu) = 1$.  The parameterisation adopted in
\cite{MARG,Block}, which is
the Fourier transform of a dipole form factor squared,  is
\be
W(b;\mu) = \frac{\mu^{2}}{96 \pi} (\mu b)^{3}K_{3}(\mu b),
\ee
where $K_{3}(x)$ is the modified Bessel function of the second kind.  
Note  the behaviour of $x^{3}K_{3}(x)$ for small $x$
\be
x^{3}K_{3}(x) = 8 + O(x^{2})
\ee
and for large $x$
\be
x^{3}K_{3}(x) = \sqrt{\frac{\pi}{2}}x^{5/2}e^{-x}(1 + O(x^{-1}).
\ee
We approximate the eikonal with an exponential 
form, 
\be
\chi_{a}(b,s) = C(s)e^{-b/b_{0}(s)},
\ee
reducing the calculation to the previous case.
 $C(s)$ is chosen to be the  opacity of the eikonal we are 
approximating, and  $b_{0}(s)$ is determined so that 
\be
\int_{0}^{\infty} b db \chi(b,s) = \int_{0}^{\infty} b db \chi_{a}(b,s),
\ee
giving
\be
C(s) = \sigma_{gg}(s)\frac{\mu^{2}_{gg}}{12\pi} + 
\sigma_{qg}(s)\frac{\mu^{2}_{qg}}{12\pi} + 
\sigma_{qq}(s)\frac{\mu^{2}_{qq}}{12\pi},
\ee
and
\be
2 \pi b_{0}^{2}(s) = \left(\sigma_{gg}(s) + \sigma_{qg}(s) + \sigma_{qq}(s)
\right)/C(s).
\ee
Note that if a single elementary collision process dominates over
other terms $b_{0}$ becomes $\sqrt {6} /\mu$, independent of $s$,
where $\mu$ is the scale
parameter of the dominant process.

Applying the analysis of the previous section, the leading large $s$ 
behaviour of the total cross section is
\be
\sigma_{t}(s) = 2\pi b_{0}^{2}(s) \ln^{2} C(s)
\ee
Block \emph{et al} \cite{Block}
show that in the large $s$ region, the eikonal is 
dominated by the gluon-gluon scattering term, which has the structure
\be
\sigma_{gg}(s) \approx \lambda 
\left(\frac{s}{m_{0}^{2}}\right)^{\epsilon} \ln\left(\frac{s}{m_{0}^{2}}\right),
\ee
where $\epsilon$ is obtained from the gluon structure function 
$f_{g}(x) = N_{g}(1-x)^{5}x^{-(1+\epsilon)}$.
The cross section again has the Heisenberg 
$\ln^{2}s$ dependence at high energies, with the universal coefficient
\be
B = 12 \pi \mu_{gg}^{-2} \epsilon^{2}.
\ee  Block \emph{et al} use the 
numerical values $\epsilon = 0.05, \mu_{gg} = 0.73 \rm{GeV}$, which 
give
$B = 0.069 \rm{mb}$, in the range of values given by the exponential 
eikonal, and rather smaller than the empirical value.
However, recent results on $\epsilon$ \cite{EPS} give a range of 
values $\epsilon \in (0.08, 0.4)$ which give $B \in (0.18, 4.42)$mb, 
covering the empirical value.
\section{Conclusion}

In the light of the results of \cite{HEA}, it is no longer sufficient 
to develop models for high energy scattering which show cross 
sections rising as $\ln^{2}s$.  It is now important to demonstrate 
that the coefficient, $B$, of the $\ln^{2}s$ term is a universal 
constant, independent of the 
hadronic reaction under consideration.

We have been able to determine the cross section exactly for both 
gaussian and exponential eikonals, and not only demonstrate the 
$\ln^{2}s$ asymptotic behaviour, but also determine its coefficient.
Each of the three eikonal models we have considered give a universal $B$.  
The numerical value of $B$ is sensitive to both the choice of eikonal 
model, and the choice of parameters within that model.  Both the 
gaussian eikonal and the QCD inspired eikonal give values of $B$ 
consistent with the observed value, and the factorisable exponential 
eikonal value is of the correct order of magnitude.  

\section*{Acknowledgements}

The authors express their thanks to President C. W. Kim for  the 
hospitality and support of the Korean Institute for Advanced Study. 

This work was supported in part by the US DOE Contract 
DE-FG-02-91ER40688-Task A, and in part by the Australian Research 
Council.

\end{document}